# Tuning orbital-selective phase transitions in a two-dimensional Hund's correlated system

Eun Kyo Ko[1,2][†], Sungsoo Hahn[1,2][†], Changhee Sohn[3], Sangmin Lee[4], Seung-Sup B. Lee[2], Byungmin Sohn[1,2], Jeong Rae Kim[1,2], Jaeseok Son[1,2], Jeongkeun Song[1,2], Youngdo Kim[1,2], Donghan Kim[1,2], Miyoung Kim[4], Choong H. Kim [1,2][*], Changyoung Kim[1,2][*], Tae Won Noh[1,2][*]

[1]*Center for Correlated Electron Systems, Institute for Basic Science (IBS), Seoul 08826, Republic of Korea*

[2]*Department of Physics and Astronomy, Seoul National University, Seoul 08826, Republic of Korea*

[3]*Department of Physics, Ulsan National Institute of Science and Technology, Ulsan, Republic of Korea*

[4]*Department of Materials Science and Engineering and Research Institute of Advanced Materials, Seoul National University, Seoul 08826, Republic of Korea*

[†]These authors contributed equally.

[*]e-mail: chkim82@snu.ac.kr, changyoung@snu.ac.kr, twnoh@snu.ac.kr

**Hund's rule coupling ($J$) has attracted much attention recently for its role in the description of the novel quantum phases of multi-orbital materials. Depending on the orbital occupancy, $J$ can lead to various intriguing phases. However, experimental confirmation of the orbital occupancy dependency has been difficult as controlling the orbital degrees of freedom normally accompanies chemical inhomogeneities. Here, we demonstrate a method to investigate the role of orbital occupancy in $J$ related phenomena without inducing inhomogeneities. By growing SrRuO$_3$ monolayers on various substrates with symmetry-preserving interlayers, we gradually tune the crystal field splitting and thus the orbital degeneracy of the Ru $t_{2g}$ orbitals. It effectively varies the orbital occupancies of two-dimensional (2D) ruthenates. Via *in-situ* angle-resolved photoemission spectroscopy, we observe a progressive metal-insulator transition (MIT). It is found that the MIT occurs with orbital differentiation: concurrent opening of a band insulating gap in the $d_{xy}$ band and a Mott gap in the $d_{xz/yz}$ bands. Our study provides an effective experimental method for investigation of orbital-selective phenomena in multi-orbital materials.**

Mott physics with strong electron correlations due to on-site Coulomb repulsion ($U$) have been a central paradigm in condensed matter[1,2]. Recently, Hund's metal, a new type of strongly correlated material, was proposed, for which Hund's rule coupling ($J$) dominantly drives the electron correlations rather than $U$[3–12]. In such a system, the roles of $J$ are highly dependent on the number of electrons/orbitals, so a variety of physical phenomena depend strongly on the orbital occupancy. In particular, rich phases with orbital differentiations have been suggested theoretically for Hund's correlated system[3,13,14]. Since the energy scale of $J$ is smaller than that of $U$ for most multi-orbital materials, the associated phase transitions can be achieved by controlling the energy scale much less than in Mott physics[3,13,14].

Despite extensive theoretical interest[3,6,7,10,13–18], direct observations of Hund-driven phase transitions have remained challenging. The $J$ value is usually determined by atomic physics, so control of its value is difficult without chemical substitution. To control orbital occupancy, earlier studies used doping and/or substitution with different chemical elements[19,20]. However, during such experiments, numerous defects are easily formed due to the random distribution of chemical elements. These inhomogeneity problems place serious limitations on experimental investigations of Hund's systems. Therefore, precise control of orbital occupancy without random chemical distribution is the key for experimentally investigating Hund-driven phase transitions.

We propose that crystal field splitting can be controlled with a suitable experimental approach for observing Hund-driven phase transitions with negligible impurity problems. In cubic perovskite oxides ($a$-lattice constant ($a$) = $c$-lattice constant ($c$)), the five $d$-orbitals of transition metal elements are split into $t_{2g}$ and $e_g$ levels due to the oxygen octahedral environment (**Fig. 1a**). When the oxygen octahedron becomes distorted, the orbitals experience further tetragonal crystal field splitting ($\Delta_t$). For instance, the three $t_{2g}$ orbitals split into $d_{xy}$ and $d_{xz/yz}$ levels if the oxygen octahedron is elongated along the in-plane directions ($c < a$). Such variation of $\Delta_t$ can tune the orbital occupancy of the $d$-orbitals without chemical doping in partially-filled $d$-electron systems.

In this study, we investigated how Hund-driven phase transitions can occur in SrRuO$_3$ (SRO) films by artificially controlling the crystal field splitting. In the bulk SRO, $\Delta_t$ is small, and the nearly degenerate $d_{xy/xz/yz}$ orbitals are partially-filled with 4 electrons (3-orbital/4-electron system). As shown in **Fig. 1b**, with an increase in $\Delta_t$, the $d_{xy}$ ($d_{xz/yz}$) band moves downward (upward), leading to a redistribution of electrons among the three $t_{2g}$ orbitals. When the energy level of the $d_{xy}$ band is lower than the Fermi energy ($E_F$), the $d_{xy}$

($d_{xz/yz}$) band becomes fully-filled (half-filled). Then, SRO will behave effectively as a 2-orbital/2-electron system. When $U/W$ is large ($W$ is the bandwidth), the $d_{xz/yz}$ bands can further experience a Mott transition[14], which is displayed schematically with the last band configuration of **Fig. 1b**.

For a comprehensive understanding of the orbital-selective phase transitions, we should pay much attention to the roles of $J$[14]. **Fig. 1c** schematically displays how the phase transitions occur in SRO for two different $J$ values. With $J = 0$, most physics will be determined solely by the $U/W$ value. On the other hand, with sizable $J$, the system should experience orbital-selective phase transitions by changing $\Delta_t$. For small $U/W$, control of $\Delta_t$ causes the system change from a Hund metallic state into another metallic state with a band insulating gap in the $d_{xy}$ band. For large $U/W$, control of $\Delta_t$ causes a phase transition into an insulating state with band ($d_{xy}$) + Mott ($d_{xz/yz}$) gaps. Note that for small $\Delta_t$, the critical $U$ value for opening the Mott gaps ($U_c$) increases when we take $J$ into account (dashed arrow (1) in **Fig. 1c**). On the other hand, with large $\Delta_t$, $U_c$ decreases with sizable $J$ (dashed arrow (2)). This variable nature of $J$ makes the Hund-driven phase transition with orbital differentiation intriguing[3,14]. Therefore, our experimental approach with control of $\Delta_t$ holds great potential for systematically exploring Hund's physics.

**Symmetry-preserving strain engineering in 2D SRO**

In this study, we observed a Hund-driven phase transition in an SRO monolayer with *in-situ* angle-resolved photoemission spectroscopy (ARPES). Thick SRO films (three-dimensional systems) did not exhibit metal-insulator transitions as shown in **Supplementary Fig. 1**. As stated earlier, the metal to band+Mott insulator transition is feasible with small control of $\Delta_t$ and large $U/W$ (red arrow in **Fig. 1c**). To magnify such effects, we investigated the SRO monolayer, which contains a two-dimensional (2D) Ru layer. When the thickness of the SRO film decreases and approaches the monolayer limit (2D), electron hopping along the z-direction diminishes[21–25]. The bandwidth of the $d_{xz/yz}$ orbitals can be further reduced by orbital-selective quantum confinement effects[26].

We used strain engineering to elongate the oxygen octahedra in SRO monolayers. We used pulsed laser deposition to grow SRO layers on five different substrates, $(LaAlO_3)_{0.3}(Sr_2TaAlO_6)_{0.7}(001)$ [LSAT(001)], $SrTiO_3(001)$ [STO(001)], $Sr_2(Al,Ga)TaO_6(001)$ [SAGT(001)], $KTaO_3(001)$ [KTO(001)], and $PrScO_3(110)$ [PSO(110)]. They are known to have (pseudocubic) lattice constants of 3.868, 3.905, 3.931, 3.989, and 4.02 Å, respectively. Since the pseudocubic lattice constant of bulk SRO is 3.923 Å, the substrates impart −1.4%, −0.5%, +0.2%, +1.7%, and +2.5% epitaxial strain on the SRO monolayers. For convenience, we have indicated compressive (tensile) strain using a minus (plus) sign. We should note that these substrates have different oxygen octahedral rotation (OOR) patterns, so each SRO monolayer grown on the above-mentioned substrates could have a different OOR pattern[27–29]. Such undesirable occurrence of the complex structural modifications could hinder our systematic investigation of Hund-driven phase transitions.

To suppress the structural complications of OOR, we developed a symmetry-preserving strain engineering technique. As shown in **Fig. 2a**, 10 unit cells (u.c.) of the $SrTiO_3$ (STO) layer were inserted between the SRO monolayer and substrates. This preserves the OOR pattern of the SRO monolayer while applying different epitaxial strains. **Fig. 2b, 2c** show low-energy electron diffraction (LEED) results for the SRO monolayers under both compressive and tensile strains at 6 K. All samples showed LEED diffraction peaks at (m + 0.5, n + 0.5) (m, n: integer), indicating the existence of OOR along the out-of-plane axis. On the other hand, no sample showed (0.5 m, 0.5 n) peaks, indicating that OOR did not occur along the in-plane axis. These LEED diffraction results indicate that all SRO monolayers on the substrates used exhibited OOR with $a^0a^0c^-$ crystal symmetry (**Supplementary Fig. 2**).

The atomic arrangements of the SRO monolayers were confirmed by scanning transmission electron microscopy (STEM). We covered the heterostructures with a 10-u.c. STO layer to protect the SRO layer from possible damage during STEM measurements. **Fig. 2d** and **Supplementary Fig. 3** show low-magnification STEM images in high-angle annular dark-field (HAADF) mode. High-magnification STEM images were acquired in the HAADF and annular bright-field (ABF) modes (**Fig. 2e, 2f**, respectively). The SRO monolayer did not show OOR along the in-plane axis, consistent with the LEED results. Although most areas showed abrupt interfaces with the SRO single layer, we observed a few regions for which thickness inhomogeneities (i.e., 0 or 2 u.c. thickness of SRO) were observed (**Supplementary Fig. 3**). These inhomogeneous regions usually occurred near step terraces[30,31], which can break the continuity of the SRO monolayer for transport measurements. Instead, we used optical and *in situ* ARPES measurements to obtain reliable area-averaged responses.

**Orbital occupancy changes**

To investigate the orbital occupancy changes, we explored the interatomic optical transitions between $Ru^{4+}$ ions via ellipsometric spectroscopy. Upon absorption of a photon, an electron can hop from one $Ru^{4+}$ ion to a nearest-neighbor ion ($d^4 + d^4 \rightarrow d^3 + d^5$ transition). The matrix element of such an interatomic transition can be significantly large due to hybridization between the Ru $d$ – O $p$ orbitals. Note that this interatomic transition can occur only between the same $t_{2g}$ orbitals due to the orbital geometry (**Fig. 3a**). Specifically, the transition from $d_{xz}$ to $d_{xy}$ (or $d_{yz}$) is not allowed, given that there is only a small overlap between the corresponding orbitals. **Fig. 3b,c** displays the allowed interatomic $d^4 + d^4 \rightarrow d^3 + d^5$ transition in SRO[32]. With a small $\Delta_t$, the three $t_{2g}$ orbitals are partially-filled with four electrons (3-orbital/4-electron). Then, interatomic transitions can occur at two photon energies ($U - J$ and $U + J$) (**Supplementary Fig. 4**)[32]. With a large $\Delta_t$, the $d_{xy}$ orbital becomes fully occupied, and the $U - J$ transition cannot occur.

We obtained an optical spectrum of the SRO monolayer under a strain of −1.4% (**Fig. 3d**). The spectrum exhibited three peaks (A–C). The A and B peaks are assigned to interatomic $d$-$d$ transitions with energy positions at $U - J$ and $U + J$, respectively. Peak C reflects a charge transfer transition from O $2p$ to Ru $t_{2g}$[33]. We fitted the spectrum (solid circles) with Lorentzian oscillators (dashed lines). The obtained peak positions were in good agreement with those previously reported for ruthenates[32,33]. The peak position difference between A and B is 2$J$; $J$ was thus ~0.6 eV[32,34]. The presence of both $U - J$ and $U + J$ peaks suggested that the compressively strained SRO monolayer contained partially-filled orbitals, which is similar to the situation for bulk SRO (**Fig. 3b**).

As tensile strain was applied, a significant spectral weight (*SW*) change occurred,

indicating strain-induced electron redistribution of the $t_{2g}$ orbitals (**Fig. 3d-f**). The change in *SW* is plotted in **Fig. 3g**. As the tensile strain was increased, the *SW* of peak A decreased and that of peak B increased. For the tensile-strained SRO monolayer (+0.2%), peak A nearly disappeared, suggesting that one of the bands (i.e., $d_{xy}$) became fully-filled. Dynamic mean-field theory (DMFT) calculations with $J = U/6$ also revealed that the filling of $d_{xy}$ ($d_{xz}/d_{yz}$) orbitals increased (decreased) with an increase in strain (**Fig. 3h**). When the strain reached +0.2%, $d_{xy}$ became fully-filled, and $d_{xz/yz}$ became half-filled. The control of orbital polarization in the Hund system can lead to an orbital-selective phase transition.

**Control of electronic structures with orbital differentiation**

The 2D metallic phase in the -1.4% strained SRO monolayer was observed by *in-situ* ARPES measurement. As the SRO monolayers have 2D atomic arrangements (**Fig. 2d-f**), the electronic structure should be similar to that of $Sr_2RuO_4$ (a well-established quasi-2D system)[35–37]. **Fig. 4a** shows the low-temperature constant energy map at $E_F$ for SRO monolayer with − 1.4% strain. The experimentally obtained Fermi surface (FS) was similar to the schematic FS of $Sr_2RuO_4$ (shown as solid lines). Therefore, we hereafter follow the notation generally used for $Sr_2RuO_4$[35–37], in which the three bands at $E_F$ are labeled α, β, and γ. The orbital characters of the α and β bands are $d_{xz/yz}$, and that of γ is $d_{xy}$.

Energy maps of the SRO monolayers show the strain-induced MIT. We measured constant energy maps for SRO monolayers at $E_F$ under epitaxial strains of −1.4, −0.5, and +0.2%. Compressively strained SRO monolayers (−1.4 and −0.5%) exhibited a nonzero density of states (DOS) at $E_F$, indicating that they are metallic (**Fig. 4a, b**). In contrast, the tensile-strained SRO monolayer (+ 0.2%) exhibited nearly zero DOS at $E_F$, indicating that it is in an insulating state (**Fig. 4c**). These strain-dependent DOSs at $E_F$ in the SRO monolayer indicate that MIT occurs at a strain of approximately +0.2%, which agrees well with the optical spectra of the SRO monolayer (**Fig. 3g, h**).

To elucidate the mechanism of the MIT, we focused on strain-dependent *E-k* dispersion along the Γ-M line (indicated by the black arrow in **Fig. 4a**) (**Fig. 4d**). Two major features depending on the epitaxial strain (-1.4, -0.5, +0.2, +1.7, and +2.5%) were observed. First, the energy level of the band near M/2 (red dashed line) moved smoothly to higher binding energies with increasing tensile strain. Second, the spectral weight (*SW*) near Γ (blue dashed line) at $E = E_F - 1.6$ eV appeared abruptly at a strain of +0.2% and was enhanced above +0.2% tensile strain.

The orbital characters of each band in the SRO monolayers were assigned by examining the constant energy maps below $E_F$ for a SRO film under a + 0.2% strain. **Fig. 5a** shows the energy distribution curve (EDC) along the Γ-M line of the SRO monolayer (+0.2%). The constant-energy maps at $E - E_F = -0.5$ eV (red dashed line) and − 1.6 eV (blue dashed line) were acquired for two experimental geometries, as shown in **Fig. 5b**. Note that the photoemission intensity for the $d_{xy}$ orbital should vary with respect to the azimuthal angle $\varphi$, whereas that for the $d_{xz/yz}$ orbital should change little with respect to $\varphi$ (**Supplementary Note**). At $E_F - 0.5$ eV, we observed a significant difference in the photoemission intensity between $\varphi = 0°$ and 45°. In particular, the intensity along the Γ-M line with $\varphi = 45°$ (**Fig. 5d**) was much lower than that with $\varphi = 0°$ (**Fig. 5c**). On the other hand, at $E_F - 1.6$ eV, the photoemission intensity at $\varphi = 0°$ and 45° showed little difference (**Fig. 5e, f**). This feature can be explained by considering the matrix element[38]. Therefore, we conclude that the orbital characteristics near the $E_F - 0.5$ eV band dominantly originated from the $d_{xy}$ orbital, and the band near $E_F -1.6$ eV dominantly originated from the $d_{xz/yz}$ orbital.

This orbital differentiation can be used to explain the evolution of the *E-k* dispersion curves in **Fig. 4d**, which confirm the rationale illustrated in **Fig. 1b**. As tensile strain was applied, the band ($d_{xy}$) near M/2 moved steadily toward higher binding energies and became a band-insulating at the MIT. Simultaneously, the band spectra near Γ ($d_{xz/yz}$) suddenly appeared close to the MIT. As $d_{xz/yz}$ became a half-filled state with tensile strain (**Fig. 3h**), the bands near Γ should be the lower Hubbard bands (LHBs). In this sense, the MIT occurred with orbital differentiation. The two kinds of gaps opened nearly simultaneously at the MIT: a band insulating gap for $d_{xy}$ and a Mott gap for $d_{xz/yz}$[34,39].

Careful analysis of the ARPES spectra also revealed how $\Delta_t$ can indeed be tuned via our symmetry-preserving strain engineering technique. The EDCs of the Γ-M cuts are shown in **Supplementary Fig. 5**. Peaks were evident near $E_F - 0.8$ to $-0.4$ eV (from $d_{xy}$) and $-1.6$ eV (from the LHBs of $d_{xz/yz}$). The peak position of $d_{xy}$ ($\zeta$) was obtained via Gaussian fitting. The $\zeta$ value (black circles) increased with tensile strain, consistent with the density functional theory (DFT) calculations without OOR along the in-plane axis (pink squares). Since the energy center position of the Hubbard bands varied little with strain, $\zeta$ should be close to $\Delta_t$. Therefore, the systematic $\zeta$ change in the ARPES indicated that $\Delta_t$ increased with an increasing in tensile strain, as expected from **Fig. 1b**.

We also tried to change the orbital occupancy by putting potassium (K) atoms on the SRO monolayer surface, which will not significantly perturb the SRO monolayer structure. K atoms will transfer electrons to the SRO monolayer. The coverage of the K layer was controlled from 0.0 monolayer (ML) to 1.0 ML. **Supplementary Fig. 6** shows K-coverage-dependent ARPES data for the SRO monolayer with +2.5% strain, which is an band ($d_{xy}$)+Mott($d_{xz/yz}$) insulator. With electron doping, the DOS for the LHBs ($d_{xz/yz}$) decreased, and the DOS near $E_F$ increased. This indicated that the correlation-induced Mott gap collapsed as the system moved away from the half-filled $d_{xz/yz}$. However, the SRO monolayer with a 1.0 ML K-coverage was still insulating (almost zero DOS at $E_F$), possibly due to strong Anderson localization effects in the 2D system[40].

**Summary & Outlook**

In summary, we demonstrated tuning of the crystal field of a SRO monolayer with symmetry-preserving strain engineering. Given the nature of 2D materials, the modulated $\Delta_t$ induced compulsory changes in orbital occupancy and dramatically resulted in an orbital-selective phase transition. Remarkably, a simultaneous MIT with orbital differentiation was observed via *in situ* ARPES measurements, in which one band became a band insulating state while the other bands opened a Mott gap. This orbital differentiation can be well explained with Hund's physics, as suggested by theoretical studies[3,6,7,12,14,16].

Our symmetry-preserving strain engineering technique can be applied to studies of various multi-orbital physics using numerous transition metal oxide monolayers[41,42]. According to recent theoretical and experimental studies[3–12], understanding Hund's physics is crucial for describing novel quantum phenomena such as unconventional superconductivity and magnetism as well as potential device applications. However, impurity problems and complex structural distortions in materials have hindered manipulation and observation of Hund's physics, such as orbital-selective Mott phases. In this context, our strategy can be used extensively for precise control of materials with simplified structures and negligible inhomogeneity problems. Thus, our work provides a way to investigate fascinating phenomena in multi-orbital systems and promote multi-orbital-based device applications.

**Methods**

**Sample preparation.** STO and SRO epitaxial layers were grown on various substrates via PLD. The targets were single-crystalline STO and polycrystalline SRO. A KrF excimer laser ($\lambda$ = 248 nm; coherent) was operated with a repetition frequency of 2 Hz. For deposition of the STO and SRO layers, the laser energies were 1.0 and 2.0 J cm$^{-2}$, respectively. The deposition temperature was 700 °C, and the oxygen pressure was 100 mTorr. The film thickness was controlled at the atomic scale by monitoring reflection high-energy electron diffraction patterns (**Supplementary Fig. 7**).

**Scanning transmission electron microscopy.** An electron-transparent STEM specimen was prepared via focused ion-beam milling (Helios 650 FIB; FEI) and further thinned via focused Ar-ion milling (NanoMill 1040; Fischione). Cross-sectional STEM images were acquired at room temperature using an instrument corrected for spherical aberration (Themis Z; Thermo Fisher Scientific Inc.) and equipped with a high-brightness Schottky-field emission gun (operating at an electron acceleration voltage of 300 kV). The semiconvergence angle of the electron probe was 17.9 mrad. The collection semiangle for ABF was 10-21 mrad.

***In situ* angle-resolved photoemission spectroscopy.** ARPES measurements were performed using home-built laboratory equipment comprising an analyzer (DA30; Scienta) and discharge lamp (Fermion instrument). He-I$\alpha$ ($hv$ = 21.2 eV) light partially polarized in a linear vertical direction (70%) was used. As all of the substrates used in this study were insulators, ARPES measurements could be hampered by charging effects arising from the sample geometry (**Fig. 2a**). Thus, we inserted an electron-absorbing layer of 4-u.c.-thick SRO between the substrate and 10 u.c. STO layer[21]. The STO layer decoupled the electronic structure of the topmost SRO monolayer from that of the inserted SRO conducting layer[21]. After growth, the SRO heterostructures were transferred to an ultrahigh vacuum (UHV; < 1.0

× 10⁻¹⁰ Torr) environment without air exposure and annealed at 570 °C for 20 min immediately before ARPES measurements.

**Low-energy electron diffraction.** All LEED data were collected using a SPECS ErLEED 1000-A. The base pressure of the UHV chamber was maintained below $8.0 \times 10^{-11}$ Torr. To match the results of the ARPES and LEED measurements, sample preparation before LEED measurements was identical to that before ARPES (preannealing of 570 °C for 20 min at $< 1.0 \times 10^{-10}$ Torr). The UHV chamber for LEED measurements was connected to the chambers used for ARPES measurements and sample growth; the sample was not exposed to air before the LEED measurements.

**Optical spectroscopy.** Optical spectroscopic characterization was performed using a spectroscopic ellipsometer (M-2000 DI; J.A. Woollam Co.). The reflectance of the bare substrate, STO (10 u.c.)/substrate, and SRO (1 u.c.)/STO (10 u.c.)/substrate were measured separately, and the optical conductivity was extracted for each layer (**Supplementary Fig. 8**). It was challenging to obtain the optical spectra of samples under higher tensile strain (+1.7 and +2.5%). The as-received KTO(001) substrate had 3~4 nm-deep surface holes. Although ARPES measurements were not seriously affected by such holes, obtaining reliable optical data from spectroscopic ellipsometry was challenging. In addition, as PSO(110) (+2.5%) single crystalline substrates have orthorhombic structures, it was quite difficult to subtract the strongly anisotropic responses.

**Density functional theory and dynamic mean-field theory calculations.** We carried out density functional theory (DFT) calculations within the Perdew-Burke-Ernzerhof exchange-correlation functional revised for solids using VASP code[43]. For the lattice constant, the experimental value of the corresponding substrate material was used for each strain configuration. We used a 600 eV plane-wave cutoff energy and 6 × 6 × 1 k-points for all DFT calculations. The internal atomic positions of the three layers closest to the surface were fully relaxed until the maximum force was below 5 meVÅ$^{-1}$. Maximally localized Wannier functions[44] for $t_{2g}$ bands were constructed for the tight-binding model to be used in DMFT calculations. We performed a single-site DMFT calculation on top of the Wannier Hamiltonian with a continuous-time QMC hybridization-expansion solver implemented in TRIQS/CTHYB.

**References**

1. Mott, N. F. Metal-Insulator Transition. *Rev. Mod. Phys.* **40**, 677–683 (1968).

2. Mott, N. F. The basis of the electron theory of metals, with special reference to the transition metals. *Proc. Phys. Soc. Sect. A* **62**, 416–422 (1949).

3. Georges, A., De Medici, L. & Mravlje, J. Strong correlations from hund's coupling. *Annu. Rev. Condens. Matter Phys.* **4**, 137–178 (2013).

4. Yin, Z. P., Haule, K. & Kotliar, G. Magnetism and charge dynamics in iron pnictides. *Nat. Phys.* **7**, 294–297 (2011).

5. Khajetoorians, A. A. *et al.* Tuning emergent magnetism in a Hund's impurity. *Nat. Nanotechnol.* **10**, 958–964 (2015).

6. Stadler, K. M., Kotliar, G., Weichselbaum, A. & von Delft, J. Hundness versus Mottness in a three-band Hubbard–Hund model: On the origin of strong correlations in Hund metals. *Ann. Phys. (N. Y).* **405**, 365–409 (2019).

7. Deng, X. *et al.* Signatures of Mottness and Hundness in archetypal correlated metals. *Nat. Commun.* **10**, 2721 (2019).

8. Yin, Z. P., Haule, K. & Kotliar, G. Kinetic frustration and the nature of the magnetic and paramagnetic states in iron pnictides and iron chalcogenides. *Nat. Mater.* **7**, 294–297 (2011).

9. Kostin, A. *et al.* Imaging orbital-selective quasiparticles in the Hund's metal state of FeSe. *Nat. Mater.* **17**, 869–874 (2018).

10. Wang, Y., Kang, C. J., Miao, H. & Kotliar, G. Hund's metal physics: From $SrNiO_2$ to

LaNiO2. *Phys. Rev. B* **102**, 161118 (2020).

11. Beugeling, W. *et al.* Topological states in multi-orbital HgTe honeycomb lattices. *Nat. Commun.* **6**, 4–10 (2015).

12. Kugler, F. B. *et al.* Strongly Correlated Materials from a Numerical Renormalization Group Perspective: How the Fermi-Liquid State of Sr2RuO4 Emerges. *Phys. Rev. Lett.* **124**, 16401 (2020).

13. De'Medici, L., Mravlje, J. & Georges, A. Janus-Faced Influence of Hund's Rule Coupling in Strongly Correlated Materials. *Phys. Rev. Lett.* **107**, 256401 (2011).

14. Huang, L., Du, L. & Dai, X. Complete phase diagram for three-band Hubbard model with orbital degeneracy lifted by crystal field splitting. *Phys. Rev. B* **86**, 035150 (2012).

15. Werner, P., Gull, E., Troyer, M. & Millis, A. J. Spin freezing transition and non-fermi-liquid self-energy in a three-orbital model. *Phys. Rev. Lett.* **101**, 166405 (2008).

16. Stadler, K. M., Kotliar, G., Lee, S. S. B., Weichselbaum, A. & Von Delft, J. Differentiating Hund from Mott physics in a three-band Hubbard-Hund model: Temperature dependence of spectral, transport, and thermodynamic properties. *Phys. Rev. B* **104**, 115107 (2021).

17. Ryee, S., Han, M. J. & Choi, S. Hund Physics Landscape of Two-Orbital Systems. *Phys. Rev. Lett.* **126**, 206401 (2021).

18. Pavarini, E., Koch, E., Richard, S. & Richard, M. The Physics of Correlated Insulators, Metals, and Superconductors. in *The Physics of Correlated Insulators, Metals, and Superconductors* vol. 7 Ch. 14 (Deutsche Nationalbibliothek, 2017).

19. Tokura, Y. & Nagaosa, N. Orbital physics in transition-metal oxides. *Science.* **288**,

462–468 (2000).

20. Lee, P. A., Nagaosa, N. & Wen, X. G. Doping a Mott insulator: Physics of high-temperature superconductivity. *Rev. Mod. Phys.* **78**, 17–85 (2006).

21. Sohn, B. *et al.* Observation of metallic electronic structure in a single-atomic-layer oxide. *Nat. Commun.* **12**, 6171 (2021).

22. King, P. D. C. *et al.* Atomic-scale control of competing electronic phases in ultrathin LaNiO 3. *Nat. Nanotechnol.* **9**, 443–447 (2014).

23. Sohn, B. *et al.* Sign-tunable anomalous Hall effect induced by two-dimensional symmetry-protected nodal structures in ferromagnetic perovskite thin films. *Nat. Mater.* **20**, 1643–1649 (2021).

24. Moon, S. J. *et al.* Dimensionality-controlled insulator-metal transition and correlated metallic state in 5d transition metal oxides $Sr_{n+1}Ir_nO_{3n+1}$ (n=1, 2, and ∞). *Phys. Rev. Lett.* **101**, 226402 (2008).

25. Valla, T. *et al.* Coherence–incoherence and dimensional crossover in layered strongly correlated metals. *Nature* **417**, 627 (2002).

26. Chang, Y. J. *et al.* Fundamental thickness limit of itinerant ferromagnetic $SrRuO_3$ thin films. *Phys. Rev. Lett.* **103**, 057201 (2009).

27. Kan, D. *et al.* Tuning magnetic anisotropy by interfacially engineering the oxygen coordination environment in a transition metal oxide. *Nat. Mater.* **15**, 432–437 (2016).

28. Rondinelli, J. M., May, S. J. & Freeland, J. W. Control of octahedral connectivity in perovskite oxide heterostructures: An emerging route to multifunctional materials discovery. *MRS Bull.* **37**, 261–270 (2012).

29. Kim, J. R. *et al.* Heteroepitaxial control of Fermi liquid, Hund metal, and Mott insulator phases in the single-atomic-layer limit. arXiv:2203.04244 (2022).

30. Boschker, H. *et al.* Ferromagnetism and Conductivity in Atomically Thin SrRuO3. *Phys. Rev. X* **9**, 011027 (2019).

31. Lagally, M. G. & Zhang, Z. Thin-film cliffhanger. *Nature* **417**, 907–909 (2002).

32. Lee, J. S. *et al.* Electron and Orbital Correlations in Ca2-xSrxRuO4 Probed by Optical Spectroscopy. *Phys. Rev. Lett.* **89**, 257402 (2002).

33. Lee, J. S. *et al.* Optical investigation of the electronic structures of Y2Ru2O7, CaRuO3, SrRuO3, and Bi2Ru2O7. *Phys. Rev. B* **64**, 245107 (2001).

34. Sutter, D. *et al.* Hallmarks of Hunds coupling in the Mott insulator Ca 2 RuO 4. *Nat. Commun.* **8**, 15176 (2017).

35. Tamai, A. *et al.* High-Resolution Photoemission on Sr2RuO4 Reveals Correlation-Enhanced Effective Spin-Orbit Coupling and Dominantly Local Self-Energies. *Phys. Rev. X* **9**, 021048 (2019).

36. Damascelli, A., Lu, D. H., K. M. Shen, N. P. Armitage, F. Ronning, D. L. Feng, C. Kim,  and Z.-X. S., Tokura, T. K. and Y. & Maeno, Z. Q. M. and Y. Fermi Surface, Surface States, and Surface Reconstruction in Sr2RuO4. *Phys. Rev. Lett.* **87**, 5194 (2000).

37. Iwasawa, H. *et al.* Interplay among Coulomb interaction, spin-orbit interaction, and multiple electron-boson interactions in Sr2RuO4. *Phys. Rev. Lett.* **105**, 226406 (2010).

38. Moser, S. An experimentalist's guide to the matrix element in angle resolved photoemission. *J. Electron Spectros. Relat. Phenomena* **214**, 29–52 (2017).

39. Riccò, S. *et al.* In situ strain tuning of the metal-insulator-transition of Ca2RuO4 in angle-resolved photoemission experiments. *Nat. Commun.* **9**, 4535 (2018).

40. Abrahams, E., Anderson, P. W., Licciardello, D. C. & Ramakrishnan, T. V. Scaling theory of localization: Absence of quantum diffusion in two dimensions. *Phys. Rev. Lett.* **42**, 673–676 (1979).

41. Kugler, F. B., Lee, S. S. B., Weichselbaum, A., Kotliar, G. & Von Delft, J. Orbital differentiation in Hund metals. *Phys. Rev. B* **100**, 115159 (2019).

42. Karp, J. *et al.* Sr2MoO4 and Sr2RuO4: Disentangling the Roles of Hund's and van Hove Physics. *Phys. Rev. Lett.* **125**, 166401 (2020).

43. Joubert, D. From ultrasoft pseudopotentials to the projector augmented-wave method. *Phys. Rev. B* **59**, 1758–1775 (1999).

44. Marzari, N. & Vanderbilt, D. Maximally localized generalized Wannier functions for composite energy bands. *Phys. Rev. B* **56**, 12847–12865 (1997).

**Acknowledgments**

All authors are grateful for the major support provided by the Research Center Program of the Institute for Basic Science of Korea (grant no. IBS-R009-D1). C.Sohn was supported by a National Research Foundation of Korea (NRF) grant funded by the Korean government (MSIT) (Creative Materials Discovery Program-No.2017M3D1A1040834) and the Ministry of Science and ICT(2020R1C1C1008734). S. Lee and M. Kim acknowledge financial support from the Korean government through the National Research Foundation (grant no. 2017R1A2B3011629). Cs-corrected STEM was performed at the Research Institute of Advanced Materials (RIAM) of Seoul National University.

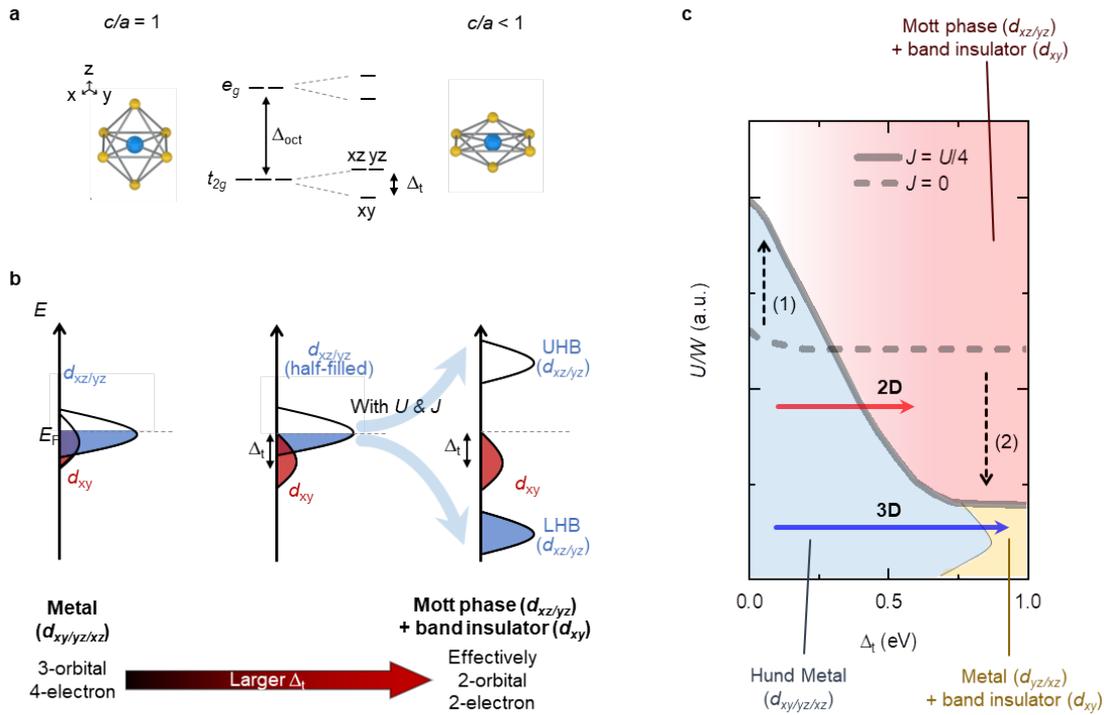

**Fig. 1 | Orbital-selective phase transition in SrRuO₃ (SRO). a,** Strain engineering for the control of the tetragonal crystal field splitting ($\Delta_t$). The energy level of the $d_{xy}$ band becomes lower than that of the $d_{xz/yz}$ bands when $c/a < 1$. **b,** Schematic of the band configurations of SRO. With a small $\Delta_t$, the partially-filled $d_{xy/xz/yz}$ is a metallic band. With a large $\Delta_t$, fully-filled $d_{xy}$ becomes band insulating, and half-filled $d_{xz/yz}$ becomes Mott insulating with sizable on-site Coulomb interaction ($U$) and Hund's rule coupling ($J$). **c,** Phase diagram of SRO depending on $U/W$ ($W$ is the bandwidth) and tetragonal crystal field splitting ($\Delta_t$) [14]. Control of $\Delta_t$ in 2D SRO can promote orbital-selective phase transition.

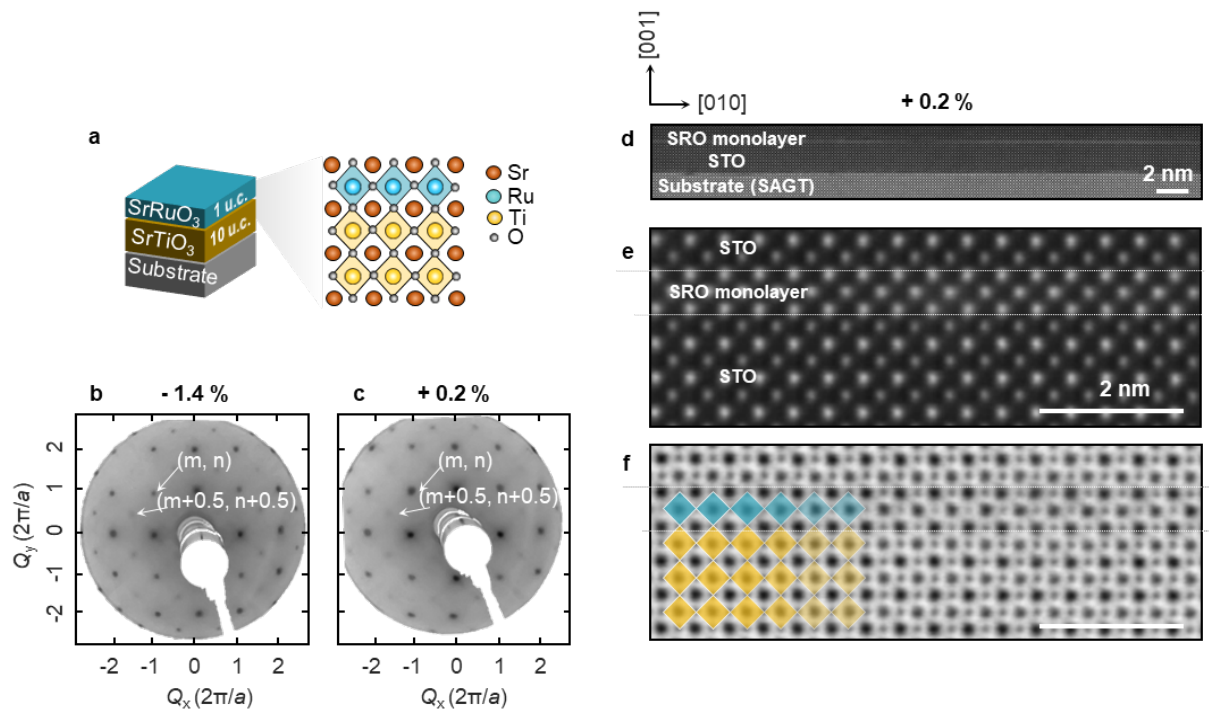

**Fig. 2 | Symmetry-preserved strained-SRO monolayers. a,** Schematic diagram of an SRO monolayer on 10 unit cells (u.c.) of SrTiO$_3$ (STO) and a substrate. The SRO monolayer has a single Ru layer sandwiched between Sr layers. **b, c,** Low-energy electron diffraction (LEED) images of SRO monolayers under −1.4 and +0.2% strain, with (m, n) and (m + 0.5, n + 0.5) peaks (m, n: integer). **d,** Structural characterization of an SRO monolayer on an STO (10 u.c.)-SAGT substrates via high-angle annular dark-field scanning transmission electron microscopy (HAADF-STEM). A 10 u.c. STO capping layer protects the SRO from damage during measurements. **e, f** HAADF (**e**) and annular bright-field (ABF) images (**f**). There is a single Ru layer with abrupt interfaces without RuO$_6$ octahedron rotation along the in-plane axis.

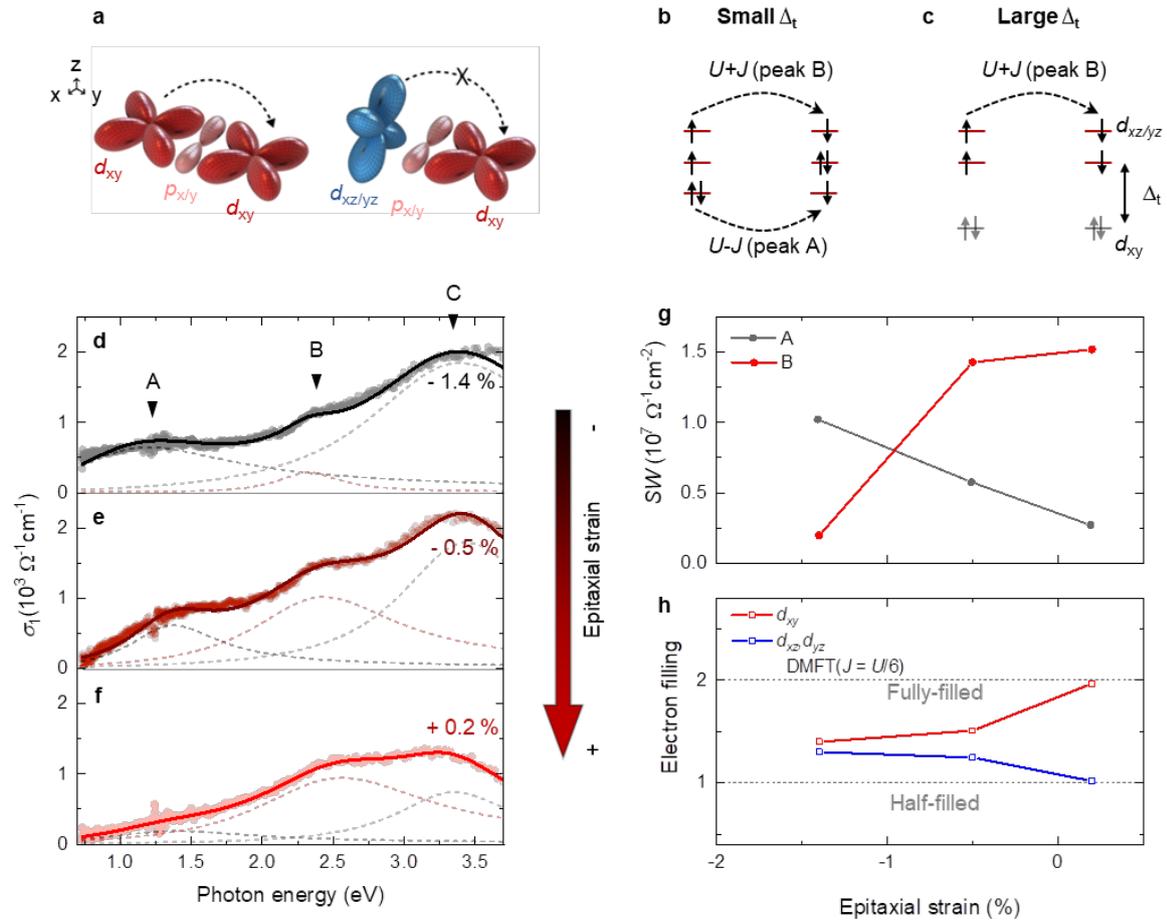

**Fig. 3 | Interatomic optical transitions and control of orbital occupancy. a,** Interatomic transitions between the *d*-orbitals, for which the matrix elements can be quite large due to hybridization between Ru *d* – O *p* orbitals. Such transitions can occur only between the same orbitals. **b,** Possible interatomic $d^4 + d^4 \rightarrow d^3 + d^5$ transitions in SRO with small $\Delta_t$. The energy costs for the two transitions differ because of *J* (i.e., $U - J$ and $U + J$). **c,** A possible interatomic $d^4 + d^4 \rightarrow d^3 + d^5$ transition in SRO with large $\Delta_t$. When $\Delta_t$ becomes large, $d_{xy}$ is fully-filled, and the $d_{xz/yz}$ orbital is half-filled. In this case, only the $U + J$ transition can occur. **d-f,** Optical spectra of SRO monolayers under various strains (−1.4, −0.5, and +0.2%) at 300 K. There are three peaks: A ($U - J$), B ($U + J$), and C (O 2$p$→Ru $t_{2g}$). **g,** Strain-dependent

spectral weight (*SW*) of peak A and peak B. The *SW* of peak A (B) decreases (increases) with an increase in strain to the plus side. **h,** Strain-dependent electron filling calculated via dynamic mean-field theory (DMFT) with $J = U/6$. When the strain reaches +0.2%, $d_{xy}$ becomes fully-filled, and $d_{xz/yz}$ becomes half-filled.

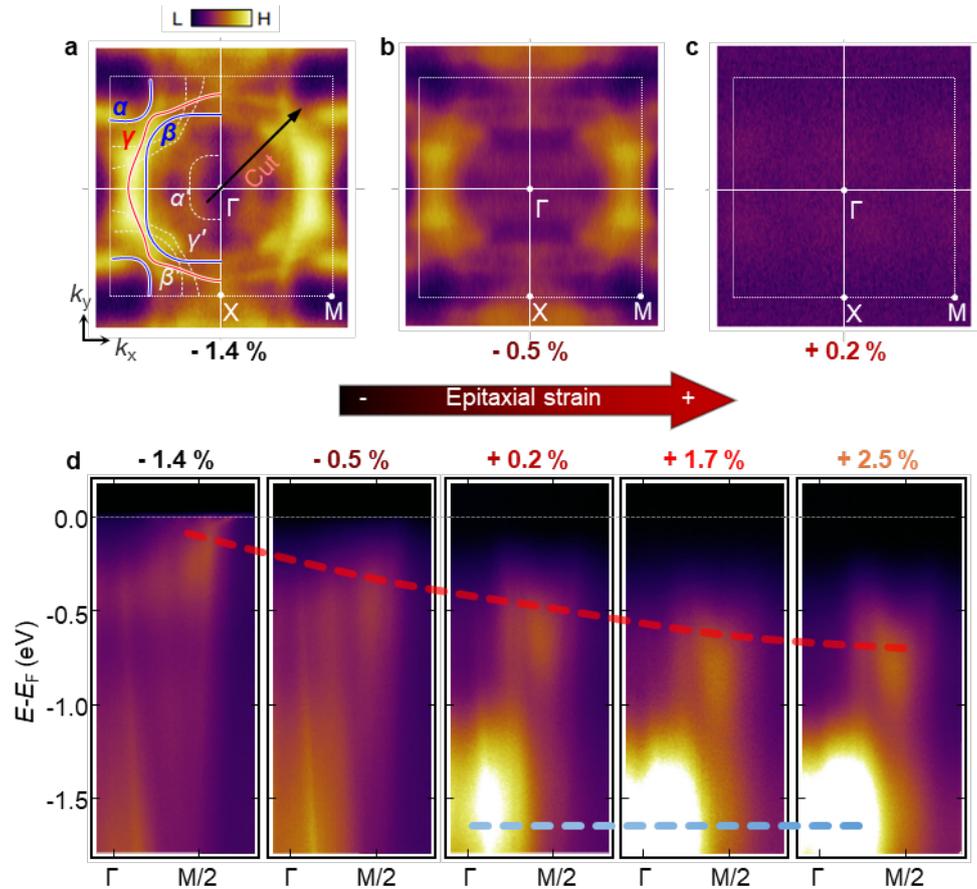

**Fig. 4 | Metal-insulator transition in SRO monolayers. a-c,** Constant energy maps at the Fermi energy ($E_F$) of SRO monolayers under −1.4, −0.5, and +0.2% strain measured at 6 K. A metal-insulator transition (MIT) is evident depending on the strain. The three bands (α, β, and γ) denoted by solid lines are assigned the notation usually used for $Sr_2RuO_4$. Folded bands attributable to the rotation of $RuO_6$ are indicated by α′, β′, and γ′ (dashed white lines). **d,** E-k dispersions along the Γ-M lines (denoted by a white arrow in **a**) under −1.4, −0.5, +0.2, +1.7, and +2.5%.

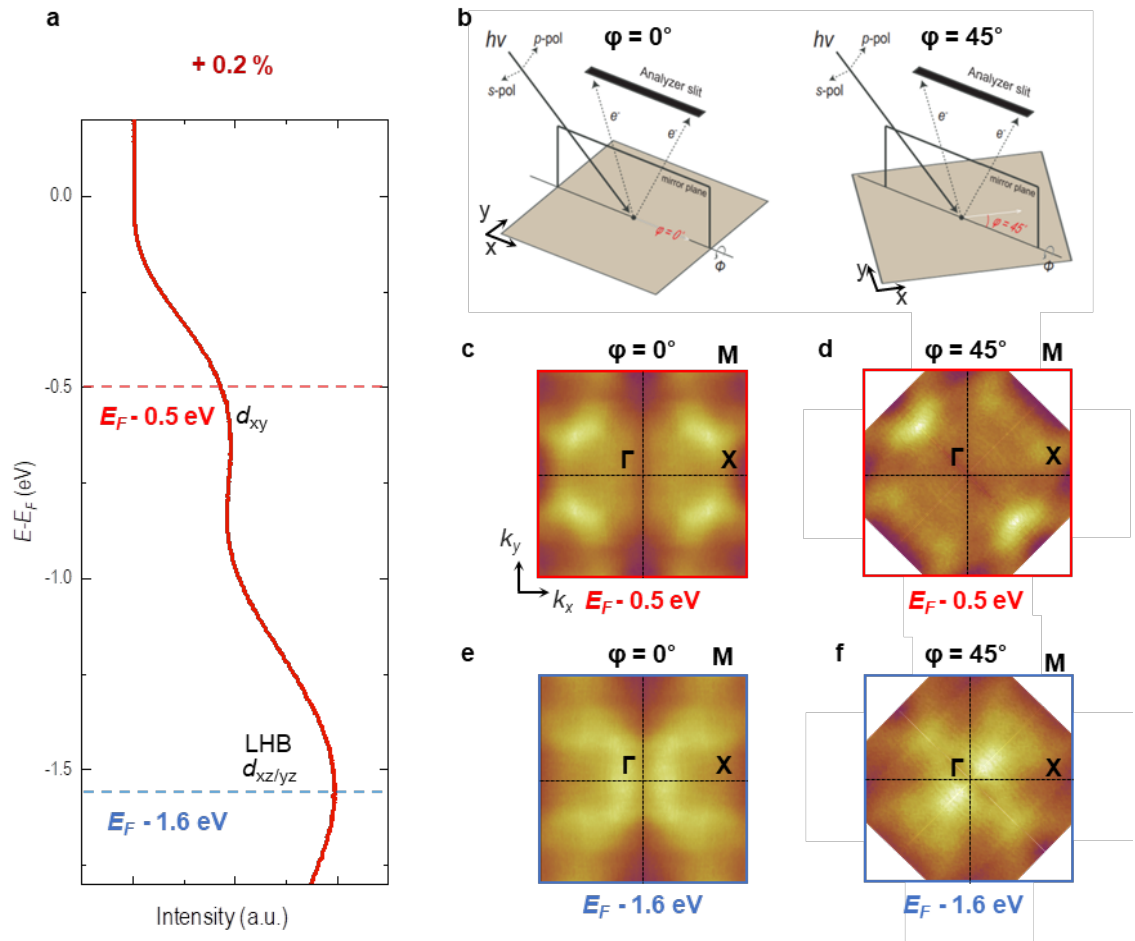

**Fig. 5 | Orbital contributions of the bands below the Fermi energy ($E_F$) in an SRO monolayer under +0.2% strain. a,** Energy distribution curve of the SRO monolayer (+0.2%) along the Γ-M line. To investigate the orbital character, we measured the constant-energy map at two azimuthal angles ($\varphi$). The maps were obtained at $E_F - 0.5$ eV (red dashed line) and $E_F - 1.6$ eV (blue dashed line) using dominantly s polarized light. **b,** Experimental geometry for $\varphi = 0°$ and 45°. **c, d,** Energy-constant maps at $E_F - 0.5$ eV with $\varphi = 0°$ (**c**) and 45° (**d**). **e, f,** Energy-constant map at $E_F - 1.6$ eV with $\varphi = 0°$ (**e**) and 45° (**f**).

# Supplementary Information

# Tuning orbital-selective phase transitions in a two-dimensional Hund's correlated system

Eun Kyo Ko[1,2†], Sungsoo Hahn[1,2†], Changhee Sohn[3], Sangmin Lee[4], Seung-Sup B. Lee[2], Byungmin Sohn[1,2], Jeong Rae Kim[1,2], Jaeseok Son[1,2], Jeongkeun Song[1,2], Youngdo Kim[1,2], Donghan Kim[1,2], Miyoung Kim[4], Choong H. Kim[1,2*], Changyoung Kim[1,2*], Tae Won Noh[1,2*]

[1]Center for Correlated Electron Systems, Institute for Basic Science (IBS), Seoul 08826, Republic of Korea

[2]Department of Physics and Astronomy, Seoul National University, Seoul 08826, Republic of Korea

[3]Department of Physics, Ulsan National Institute of Science and Technology, Ulsan, Republic of Korea

[4]Department of Materials Science and Engineering and Research Institute of Advanced Materials, Seoul National University, Seoul 08826, Republic of Korea

[†]These authors contributed equally.

[*]e-mail: chkim82@snu.ac.kr, changyoung@snu.ac.kr, twnoh@snu.ac.kr

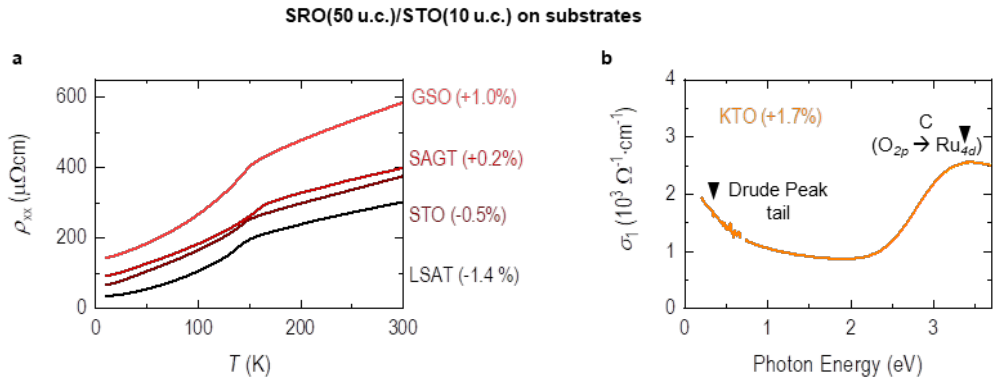

**Fig. S1 | Metallic states of thick SrRuO₃ (SRO) films. a,** Temperature-dependent longitudinal resistivity ($\rho_{xx}$) of a 50unit cell (u.c.) SRO/10 u.c. SrTiO$_3$(STO) film on various substrates. Compressive (tensile) strain is indicated by the minus (plus) sign. The strain is calculated by reference to the lattice constant of bulk SRO. All samples behave as metals. **b,** Optical spectrum of SRO (50 u.c.) deposited on a KTaO$_3$ (KTO) substrate (+1.7%). Measurement of the transport properties of SRO (50 u.c.) on the KTO substrate was challenging because the large tensile strain in thick films created cracks[1], rendering transport path discontinuities. Instead, we measured the optical spectra (in the infrared and ultraviolet regions) via ellipsometry. A Drude peak for the thick film under +1.7% strain is apparent, indicating the metallic state.

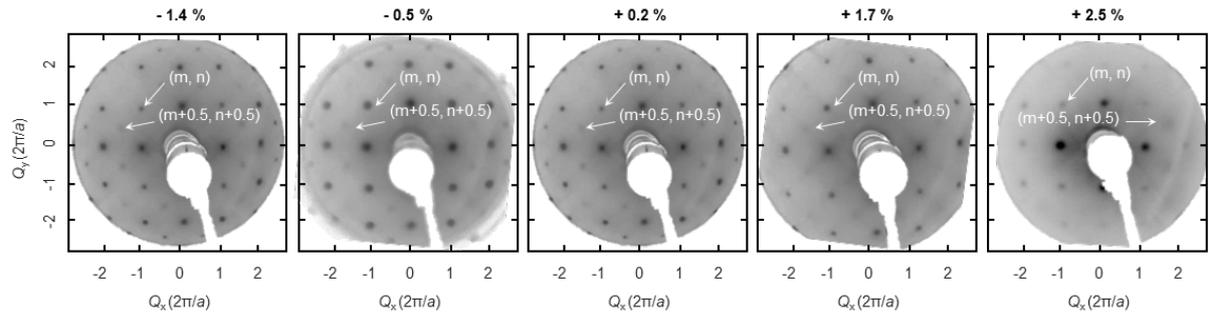

**Fig. S2 | Structural characterization of a SRO monolayer via low-energy electron diffraction (LEED).** For any symmetry, there can be (m, n) LEED peaks (m, n: integers). If there is $\sqrt{2} \times \sqrt{2}$ reconstruction due to the rotation of $RuO_6$ (along the out-of-plane), (m + 0.5, n + 0.5) peaks should appear. If there is $2 \times 1$ reconstruction due to tilting of $RuO_6$ (along the in-plane), (0.5 m, 0.5 n) peaks should appear. The LEED images of our SRO monolayers on various substrates were captured at 200 eV and 6 K. All samples exhibited (m + 0.5, n + 0.5) peaks without (0.5 m, 0.5 n) peaks. Therefore, we can conclude that the structural symmetries of the SRO monolayers are identical, independent of the applied epitaxial strain and substrates. The preserved symmetry is attributable to the inserted STO layers[2].

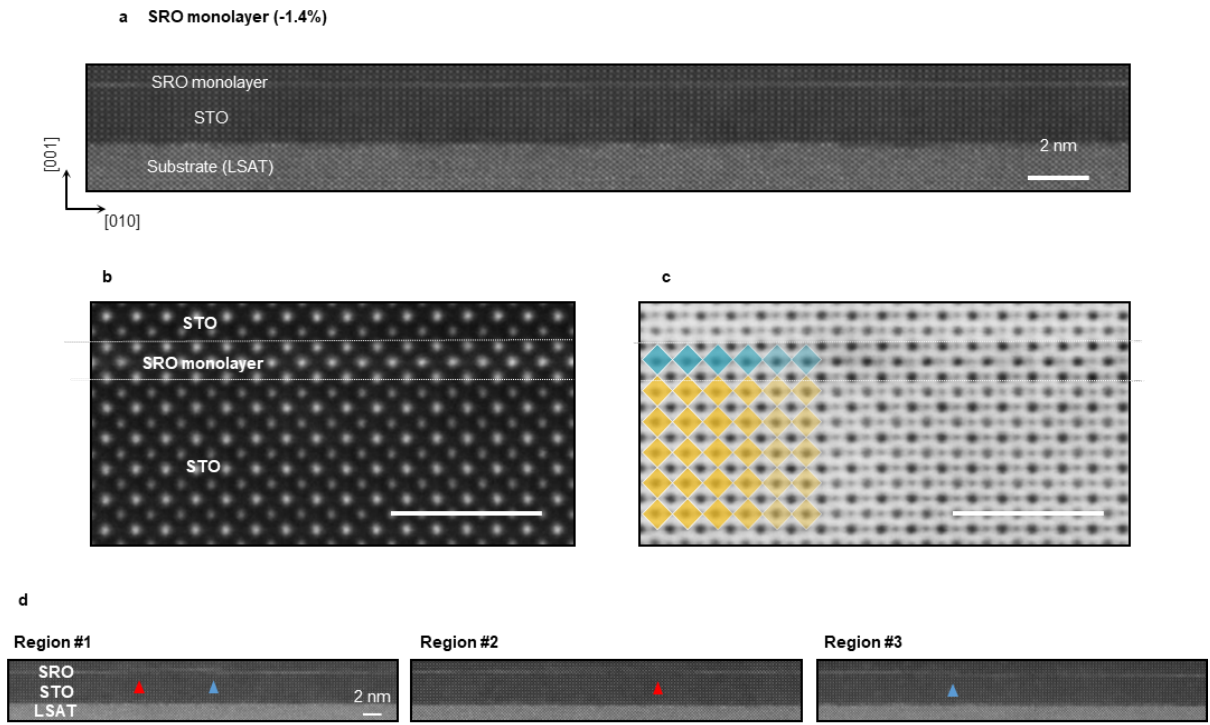

**Fig. S3 | Scanning transmission electron microscopy (STEM) images of SRO on LSAT(001). a,** Structural characterization of an SRO monolayer on a STO (10 u.c.)-LSAT substrate via high-angle annular dark-field (HAADF) mode. A 10 u.c. STO cap protects the SRO from damage during measurements. **b, c** HAADF (**b**) and annular bright-field images (**c**); a single Ru layer with abrupt interfaces is revealed, without $RuO_6$ octahedron tilting. **d,** HAADF-STEM images of different regions in SRO on LSAT(001). Most regions have atomically sharp interfaces with the SRO monolayer, as shown in **Fig. 2** (main text). However, a few regions exhibit thickness inhomogeneities. The red arrows indicate regions with discontinuous Ru layers (0 u.c. SRO). The blue arrows indicate regions with Ru double layers (2 u.c. SRO).

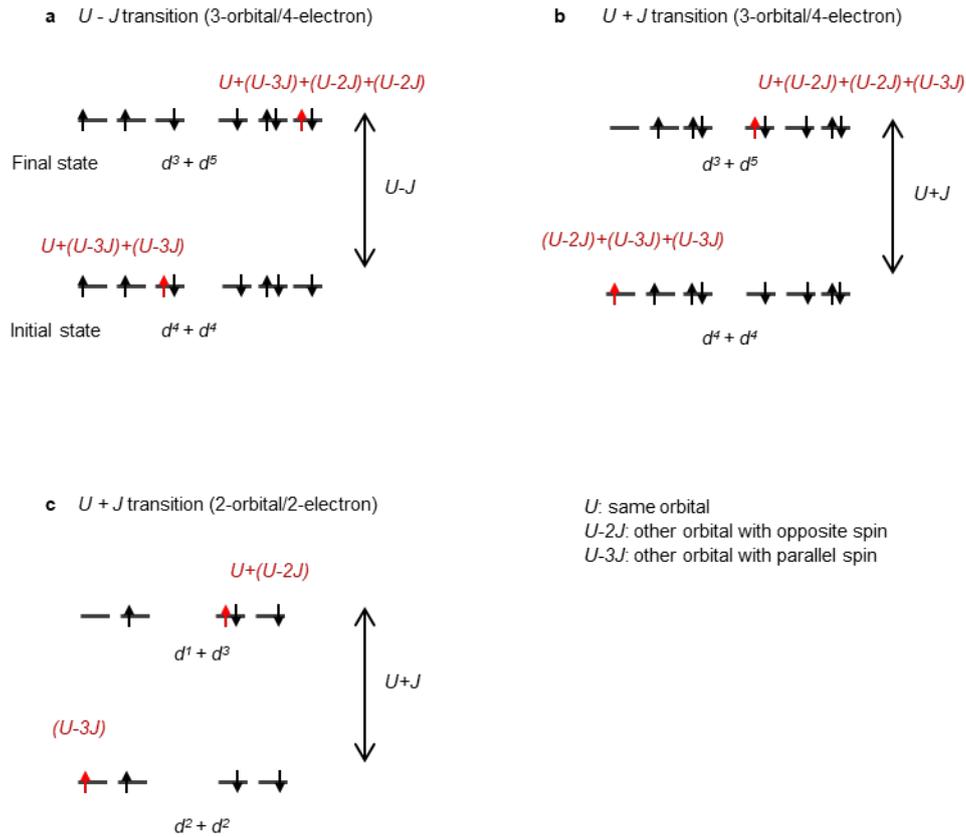

**Fig. S4 | Interatomic *d-d* transitions in ruthenates. a, b,** Interatomic *d-d* transition with an energy cost of $U - J$ (**a**) and $U + J$ (**b**) in three-orbital/four-electron (hereafter, 3-orbital/4-electron) systems. **c,** Interatomic *d-d* transition with an energy cost of $U + J$ in two-orbital/two-electron (hereinafter, 2-orbital/2-electron) systems.

The interactions between electrons in the same orbital are attributed to $U$. When the rotational symmetry of the Hamiltonian for $t_{2g}$ orbitals is considered, the interactions between electrons in different orbitals (i.e., with opposite spin) can be described as $U - 2J$, where $J$ is the Hund's rule coupling. The interactions between electrons in different orbitals with the same spin can be described as $U - 3J$. The energy difference between the initial ($d^4 + d^4$) and final ($d^3 + d^5$) states is the energy cost of the transition.

Consider the possible $d^4 + d^4 \rightarrow d^3 + d^5$ transition for a material with a (3-orbital/4-electron) system. Here, we consider only cases with paramagnetic or antiferromagnetic ordering because our SRO monolayer does not exhibit ferromagnetic ordering. Due to the Pauli exclusion principle, electrons with the same spin cannot occupy the same orbital state. Thus, there should be two kinds of interatomic transitions. Given $J$, they have different energies: $U - J$ and $U + J$.

When $\Delta_t$ is large, $d_{xy}$ is fully occupied; this effectively represents a (2-orbital/2-electron) system. In such a case, only the $U + J$ transition occurs because all involved orbitals are half filled (under the influence of $J$).

The results shown in **Fig. 3** (main text) confirm the role played by $J$ in electron distributions. Assuming that there is no $J$, in an effective 2-orbital/2-electron system, the electron configurations may be (↑↓, 0), (0, ↑↓), (↑, ↑), etc. In suchcases, fully-filled orbitals whose nearest-neighbor ion has half-filled (or empty) orbitals will induce peak A ($U - J$). However, as shown in **Fig. 3**, under strain, the $U - J$ transition disappeared from our SRO, indicating the existence of strong $J$ effects. In other words, following Hund's rule, the two electrons are distributed equally in the two orbitals, preferring the high-spin configuration.

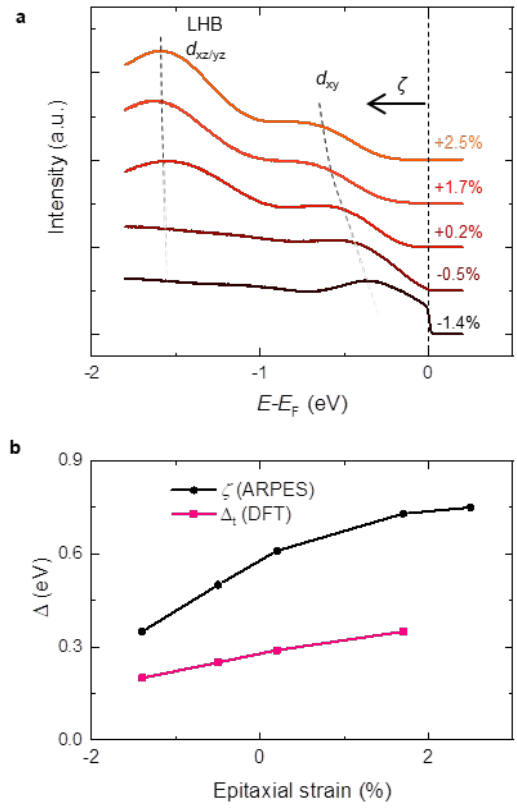

**Fig. S5 | MIT with $\Delta_t$ modulation. a,** EDCs of the Γ-M cuts or SRO monolayers under −1.4, −0.5, +0.2, +1.7, and +2.5%. The band near M/2 has a $d_{xy}$ character, whereas that near Γ has a $d_{xz/yz}$ character. **b,** Strain-dependent positions of the $d_{xy}$ bands (ζ) and $\Delta_t$ calculated via density functional theory (DFT).

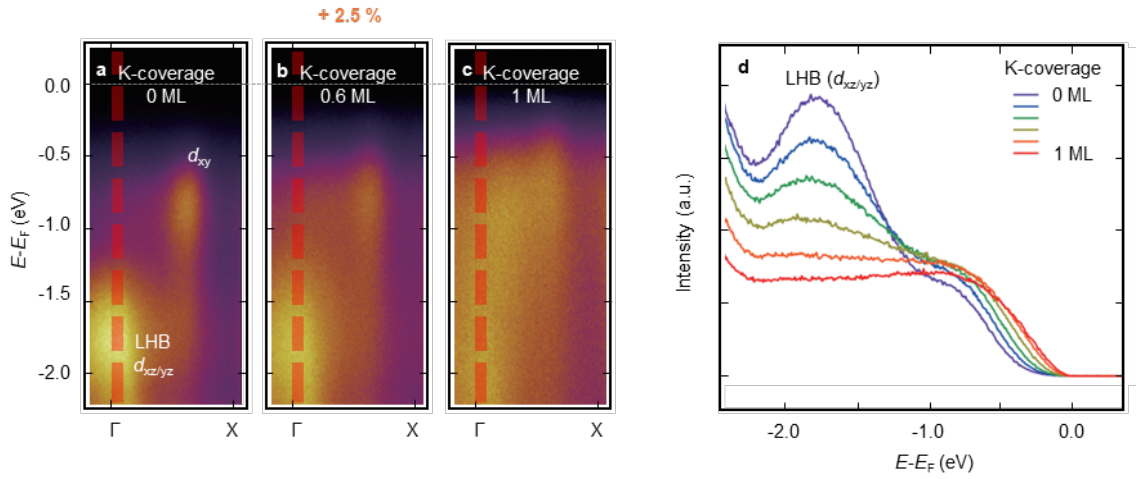

**Fig. S6 | Doping studies of the effective (2-orbital/2-electron) system**. **a-c,** *E-k* dispersions of an SRO monolayer with +2.5% strain along the Γ-X line. The K-coverage ranged from 0.0 monolayer (ML) to 1.0 ML. **d,** EDCs near the Γ point (red dashed lines in **a–c**) over the range of $0.0 \text{ Å}^{-1} \leq k_x \leq 0.1 \text{ Å}^{-1}$ and $k_y = 0.0 \text{ Å}^{-1}$.

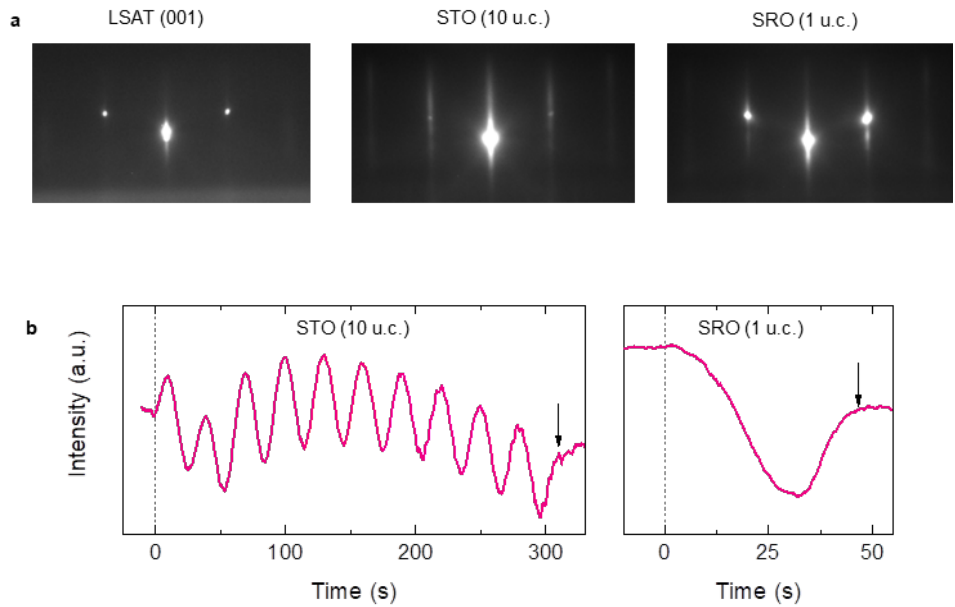

**Fig. S7 | Reflection high-energy electron diffraction (RHEED) images taken during deposition. a,** RHEED patterns of the LSAT(001) substrate, STO (10 u.c.) film, and a monolayer SRO film along the $[100]_{pseudo-cubic}$ direction. **b,** RHEED intensities. On the left, a 10 u.c. STO layer is deposited on an LSAT substrate in layer-by-layer growth mode. On the right, 1 u.c. of $SrRuO_3$ (SRO) is deposited on a STO (10 u.c.)-LSAT substrate.

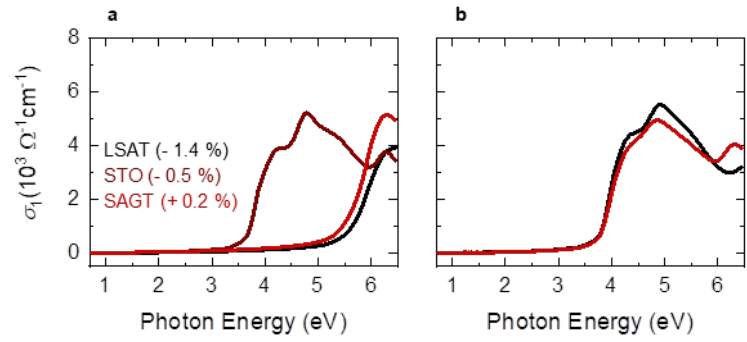

**Fig. S8| Optical spectra of substrates and STO (10 u.c.) layers. a,** Optical spectroscopy results for LSAT, STO, and SAGT substrates. **b,** Optical spectroscopy results for STO (10 u.c.) layers deposited on LSAT and SAGT substrates. Band-insulating gaps near 3.7 eV are evident, consistent with bulk STO.

**Supplementary Note: azimuthal angle dependent APRES intensity**

Note that the photoemission intensity strongly depends on the matrix element, i.e., $\langle f|\boldsymbol{A}\cdot\boldsymbol{p}|i\rangle$. $|i\rangle$ and $|f\rangle$ are the initial state and final state of the photoemitted electron, respectively. $\boldsymbol{A}$ and $\boldsymbol{p}$ are the photon polarization vector and electron momentum operator, respectively. The symmetry of $|f\rangle$ varies by the experimental geometry (such as φ), whereas that of $|i\rangle$ is fixed by the crystal lattice. Let us consider the matrix element for the $d_{xy}$ orbital. The symmetry of $|i\rangle$ is fixed as even in our experimental geometry. On the other hand, the symmetry of $|f\rangle$ in $d_{xy}$ orbital is odd with φ = 0°, while it is even with φ = 45°. With s polarized light, the matrix element for the $d_{xy}$ at φ = 45° should vanish. In other words, the ARPES intensity of $d_{xy}$ with φ = 45° should be smaller than that with φ = 0°.

At an $E_F$ of –0.5 eV, the intensity along Γ-M with φ = 45° is significantly lower than that with φ = 0°. Therefore, the orbital characters near an $E_F$ of –0.5 eV originate from $d_{xy}$ dominantly. On the other hand, at an $E_F$ of –1.6 eV, the intensity along Γ-M at φ = 45° does not decrease compared to that at φ = 0°. This suggests a contribution by the $d_{xz/yz}$ orbital dominantly.

# References

1. Wang, H. *et al.* Direct Observation of Huge Flexoelectric Polarization around Crack Tips. *Nano Lett.* **20**, 88–94 (2020).

2. Kim, J. R. *et al.* Heteroepitaxial control of Fermi liquid, Hund metal, and Mott insulator phases in the single-atomic-layer limit. arXiv:2203.04244 (2022).

3. Moser, S. An experimentalist's guide to the matrix element in angle resolved photoemission. *J. Electron Spectros. Relat. Phenomena* **214**, 29–52 (2017).

4. Hu, B. *et al.* Surface and bulk structural properties of single-crystalline Sr 3Ru2O7. *Phys. Rev. B - Condens. Matter Mater. Phys.* **81**, 184104 (2010).